\documentclass[ALICE,manyauthors]{cernphprep}
\usepackage[comma,square,numbers,sort&compress]{natbib}

\usepackage{lineno}
\usepackage{xspace}
\usepackage{hyperref}
\usepackage[T1]{fontenc}
\usepackage{color}
\usepackage{orcidlink}

\newcommand{\pp}           {pp\xspace}

\newcommand{\PbPb}         {\mbox{Pb--Pb}\xspace}

\newcommand{\pPb}          {\mbox{p--Pb}\xspace}
\newcommand{\AuAu}         {\mbox{Au--Au}\xspace}
\newcommand{\UU}           {\mbox{U--U}\xspace}

\newcommand{\snn}          {\ensuremath{\sqrt{s_{\mathrm{NN}}}}\xspace}
\newcommand{\pt}           {\ensuremath{p_{\rm T}}\xspace}

\newcommand{\etarange}[1]  {\ensuremath{\left | \eta \right | < #1}\xspace}

\newcommand{\Npart}        {\ensuremath{\langle N_\mathrm{part} \rangle}\xspace}

\newcommand{\nineH}        {$\sqrt{s}~=~0.9$~Te\kern-.1emV\xspace}
\newcommand{\seven}        {$\sqrt{s}~=~7$~Te\kern-.1emV\xspace}
\newcommand{\twoH}         {$\sqrt{s}~=~0.2$~Te\kern-.1emV\xspace}
\newcommand{\twosevensix}  {$\sqrt{s}~=~2.76$~Te\kern-.1emV\xspace}
\newcommand{\five}         {$\sqrt{s}~=~5.02$~Te\kern-.1emV\xspace}
\newcommand{\twosevensixnn}{$\sqrt{s_{\mathrm{NN}}}~=~2.76$~Te\kern-.1emV\xspace}
\newcommand{\fivenn}       {$\sqrt{s_{\mathrm{NN}}}~=~5.02$~Te\kern-.1emV\xspace}
\newcommand{\thirteen}     {$\sqrt{s}~=~13$~Te\kern-.1emV\xspace}
\newcommand{\LT}           {L{\'e}vy--Tsallis\xspace}
\newcommand{\GeVc}         {Ge\kern-.1emV/$c$\xspace}
\newcommand{\MeVc}         {Me\kern-.1emV/$c$\xspace}
\newcommand{\TeV}          {Te\kern-.1emV\xspace}
\newcommand{\GeV}          {Ge\kern-.1emV\xspace}
\newcommand{\MeV}          {Me\kern-.1emV\xspace}
\newcommand{\GeVmass}      {Ge\kern-.2emV/$c^2$\xspace}
\newcommand{\MeVmass}      {Me\kern-.2emV/$c^2$\xspace}

\newcommand{\ZDC}          {\rm{ZDC}\xspace}

\newcommand{\SPD}          {\rm{SPD}\xspace}

\newcommand{\VZERO}        {\rm{V0}\xspace}

\newcommand{\acceff}[1][]{\ensuremath{A\epsilon^{#1}}\xspace}
\newcommand{\br}{\ensuremath{\mathrm{BR}_{\jpsi\to\mu^+\mu^-}}\xspace}
\newcommand{\dxdpt}[1]{\ensuremath{\frac{\mathrm{d}#1}{\mathrm{d}\pt}}\xspace}

\newcommand{\jpsi}{\mbox{J\kern-0.05em /\kern-0.05em$\psi$}\xspace}
\newcommand{\nmb}[1][]{\ensuremath{N_{\rm MB}^{#1}}\xspace}
\newcommand{\nraw}[1][]{\ensuremath{N_{\jpsi}^{#1}}\xspace}
\newcommand{\raa}[1][]{\ensuremath{R_{\rm AA}^{#1}}\xspace}
\newcommand{\sigmapp}{\ensuremath{\sigma_{\rm pp}}\xspace}
\newcommand{\taa}[1][]{\ensuremath{\langle T_{\mathrm{AA}}^{#1}\rangle}\xspace}
\newcommand{\avpt}{\ensuremath{\langle \pt \rangle}\xspace}

\newcommand{\hh}{\ensuremath{\mathrm{h}}}

\newcommand{\ffi}{\ensuremath{f_{\mathrm{I}}}\xspace}
\newcommand{\ffd}{\ensuremath{f_{\mathrm{D}}}\xspace}
\begin{document}

\begin{titlepage}
\PHyear{2022}       
\PHnumber{071}      
\PHdate{29 March}  

\title{Photoproduction of low-\pt \jpsi from peripheral to central \PbPb collisions at 5.02 TeV}
\ShortTitle{Photoproduction of low-\pt \jpsi in \PbPb collisions at 5.02 TeV}   

\Collaboration{ALICE Collaboration\thanks{See Appendix~\ref{app:collab} for the list of collaboration members}}
\ShortAuthor{ALICE Collaboration} 

\begin{abstract}
An excess of \jpsi yield at very low transverse momentum ($\pt < 0.3$~\GeVc), originating from coherent photoproduction, is observed in peripheral and semicentral hadronic \PbPb collisions at a center-of-mass energy per nucleon pair of \fivenn.
The measurement is performed with the ALICE detector via the dimuon decay channel at forward rapidity ($2.5<y<4$).
The nuclear modification factor at very low \pt and the coherent photoproduction cross section are measured as a function of centrality down to the 10\% most central collisions.
These results extend the previous study at \twosevensixnn, confirming the clear excess over hadronic production in the \pt range $0-0.3$~\GeVc and the centrality range 70--90\%, and establishing an excess with a significance greater than 5$\sigma$ also in the 50--70\% and 30--50\% centrality ranges.
The results are compared with earlier measurements at \twosevensixnn and with different theoretical predictions aiming at describing how coherent photoproduction occurs in hadronic interactions with nuclear overlap.
\end{abstract}
\end{titlepage}

\setcounter{page}{2} 


Diffractive photoproduction of \jpsi mesons in nucleus--nucleus collisions is sensitive to  the nuclear gluon distributions at low Bjorken-$x$, in the range $x$ $\sim$ 10$^{-5}$ to 10$^{-2}$ at LHC energies, where they are still poorly constrained~\cite{Eskola:2016oht,Guzey:2013xba,Guzey:2020ntc}.
This process was extensively studied in nuclear collisions with impact parameters larger than twice the nuclear radius, known as ultra-peripheral collisions (UPCs)~\cite{ALICE:2012yye,Abbas:2013oua,Khachatryan:2016qhq,ALICE:2019tqa,Acharya:2021bnz,ALICE:2021gpt}.
In UPCs, hadronic interactions are strongly suppressed providing a clean experimental environment to study photon-induced processes.

Photonuclear reactions are produced by the strong electromagnetic field generated by ultra-relativistic ions, which can be treated as a flux of quasi-real photons.
At leading order in perturbative quantum chromodynamics (pQCD), the photon fluctuates into a quark--antiquark pair (a color dipole)~\cite{Ryskin:1992ui}, which probes the gluon distribution of the target via the exchange of two gluons in a singlet color state, with the dipole finally recombining into a vector meson (VM)~\cite{Klein:2019qfb,Baltz:2007kq}.
The diffractive VM photoproduction on nuclei can be either coherent or incoherent. 
In the coherent interaction, the photon couples with the nucleus as a whole, leaving it intact.
The produced VM is characterized by a very low average transverse momentum (\avpt $\approx$ 60~\MeVc).
In incoherent photoproduction the photon couples to a single nucleon which leads to the breakup of the nucleus.
In this case a VM with larger average transverse momentum (\avpt $\approx$ 500~\MeVc) is produced.

In nuclear collisions with impact parameters smaller than the sum of the radii of the colliding nuclei, production from hadronic interactions becomes the dominant contribution to the \jpsi yield.
Hadroproduction of \jpsi mesons in \PbPb collisions is a long-standing probe of the quark--gluon plasma (QGP), a state of strongly-interacting matter characterized by quark and gluon degrees of freedom predicted by QCD to exist at high temperature and energy density.
Charmonium production is affected by the QGP, and their measured yields~\cite{Abelev:2013ila, Adam:2015isa, Acharya:2019iur,Acharya:2019lkh} are explained as an interplay between suppression due to color screening~\cite{Matsui:1986dk} and recombination of charm quarks~\cite{BraunMunzinger:2000px,Thews:2000rj, Andronic:2007bi}.
Finally, the charmonium yield can also be influenced by cold nuclear matter effects (CNM), which can be studied independently in \pPb collisions~\cite{ALICE:2018mml, ALICE:2018szk, Sirunyan:2017mzd, Aaij:2017cqq}.

The ALICE Collaboration reported the presence of an unexpectedly large \jpsi yield at very low \pt in peripheral \PbPb collisions at a center-of-mass energy per nucleon pair of \twosevensixnn~\cite{Adam:2015gba}, which could not be explained by any combination of suppression, regeneration, and CNM effects~\cite{Shi:2017qep}.
Coherent photoproduction of \jpsi in \PbPb collisions with nuclear overlap was proposed as a plausible mechanism to explain this observation~\cite{Adam:2015gba}.
A similar low-\pt \jpsi excess was later measured by the STAR Collaboration at RHIC in \AuAu collisions at $\snn = 200$~\GeV and \UU collisions at $\snn = 193$~\GeV~\cite{STAR:2019yox}.
The STAR measurement of the $t$-dependence (Mandelstam variable, $t \approx -\pt^2$ for large \snn) of the excess showed a strong similarity with the one measured in UPCs, also pointing to coherent photoproduction as the origin of the excess.
Similar conclusions can be drawn from the recent measurement of the \jpsi yields at low \pt in \PbPb collisions at \fivenn by the LHCb Collaboration~\cite{LHCb:2021hoq}.
In addition, dilepton pairs with characteristics compatible with  photoproduction were observed in non-UPC heavy-ion collisions by the ATLAS, STAR, and ALICE experiments~\cite{ATLAS:2018pfw, STAR:2018ldd, ALICE:2022hvk}.

The concept of coherent photoproduction in a hadronic environment raises several theoretical challenges.
For example, how can the coherence condition survive in the photon--nucleus interaction if the latter is broken up during the hadronic collision?
Do only the (non-interacting) spectator nucleons participate in the coherent process?
To what extent is the photon-nucleus cross section modified by target nucleons undergoing hadronic interactions and losing energy before the photoproduction occurs?
How is the yield of the photoproduced \jpsi mesons, characterized by low transverse momenta, affected by interactions with the formed and fast-expanding QGP medium? 
The measurements mentioned above triggered novel theoretical developments~\cite{GayDucati:2018who,Cepila:2017nef,Zha:2017jch,Klusek-Gawenda:2015hja,Shi:2017qep,Jenkovszky:2022wcw} based on calculations for UPCs in which the nuclear photoproduction cross section of a VM is usually computed as the product of a quasi-real photon flux with the photon-nucleus cross section corresponding to the $\gamma \rm A \rightarrow VM + A$ interaction, where $\gamma$ is the photon and A is the nucleus.
For collisions with nuclear overlap, in all considered models, an effective photon flux is introduced to take into account the geometrical constraints of a given impact-parameter range.
Depending on the model, the photon-nucleus cross section is sometimes also modified to account for the effective size of nuclear fragments participating in the coherent process~\cite{GayDucati:2018who,Zha:2017jch}.
Calculations from Ref.~\cite{Zha:2017jch} highlight the interest of measuring the cross section (and additionally its transverse momentum dependence) of the \jpsi excess towards more central collisions in order to probe possible changes of the effective size of the coherently interacting volume.
Additionally, it was suggested that the measurement of the \jpsi coherent photoproduction in UPCs and in peripheral collisions in the same rapidity range at forward rapidity can be used to extract the coherent photon-nucleus cross section in two different Bjorken-$x$ regions, below $10^{-4}$ and above $10^{-2}$ at LHC energies~\cite{Contreras:2016pkc}.

In this Letter, the measurement of the \jpsi nuclear modification factor and the coherent photoproduction at low \pt at forward rapidity in \PbPb collisions at \fivenn are presented as a function of collision centrality.
The measurement uses a pp reference at the same energy that is described in Ref.~\cite{ALICE:2021qlw}.
The larger data set compared to the one at \twosevensixnn~\cite{Adam:2015gba} allows for the first time the observation of a significant excess in the 50--70\% and 30--50\% centrality intervals.
Assuming that the observed excess originates from coherent \jpsi photoproduction, the corresponding cross section is extracted as a function of the collision centrality.
For centrality intervals where no significant excess could be measured, an upper limit on the cross section is reported.

The ALICE detector and its performance are described in detail in Refs.~\cite{Aamodt:2008zz,Abelev:2014ffa}. 
In this analysis, the \jpsi production is measured at forward rapidity ($2.5<y<4$) and down to $\pt = 0$ in the dimuon decay channel with the forward muon spectrometer, consisting of a tracking system placed downstream of a front absorber of composite material, and a trigger system placed downstream of a muon filter made of iron.
The interaction vertex is determined with the Silicon Pixel Detector (\SPD), which consists of the two innermost layers of the Inner Tracking System in the central barrel.
The first and second innermost layers cover the pseudorapidity ranges \etarange{2} and \etarange{1.4}, respectively.
The \VZERO detector, consisting of two scintillator hodoscopes placed on both sides of the interaction point and covering the pseudorapidity range $2.8<\eta<5.1$ and $-3.7<\eta<-1.7$, is used for triggering, beam--gas background rejection and determination of the collision centrality, which is evaluated by fitting the signal amplitude distribution in the \VZERO as described in Ref.~\cite{Abelev:2013qoq}.
The Zero Degree Calorimeters ({\ZDC}s) are placed on both sides of the interaction point along the beam direction at a distance of 112.5~m from it and measure the spectator protons and neutrons.
The requirement of a minimum energy deposited in the two neutron calorimeters, corresponding to the expected signal from one spectator neutron, and the combined use of the \VZERO and \ZDC timing information, suppresses the background induced by electromagnetic dissociation processes~\cite{ALICE:2012aa}.

The data sample considered in this analysis, collected in 2015 and 2018, consists of events where two opposite sign muons are detected in the trigger system of the muon spectrometer, each with a \pt above the trigger threshold of 1~\GeVc, in coincidence with a minimum-bias (MB) trigger.
The latter is defined by the coincidence of a signal in both arrays of the \VZERO detector.
Events are selected in the 0--90\% centrality interval, where the MB trigger is fully efficient.
The data sample used for this analysis amounts to $4\times 10^8$ triggered \PbPb collisions, corresponding to an integrated luminosity of $756 \pm 19$~$\mu \rm b^{-1}$~\cite{ALICE-PUBLIC-2021-001}, where the uncertainty is systematic (the statistical one being negligible).

 \jpsi candidates are formed by combining pairs of opposite-sign (OS) muon tracks reconstructed in the geometrical acceptance of the muon spectrometer ($-4<\eta<-2.5$).
 The muon identification is ensured by requiring that each track reconstructed in the tracking chambers matches a track segment in the trigger system.
 The single-muon and dimuon selection criteria are the same as the ones used in previous analyses~\cite{Adam:2015isa,Adam:2015gba}.
 The raw number of \jpsi is extracted in five centrality classes (0--10\%, 10--30\%, 30--50\%, 50--70\% and 70--90\%) and two \pt ranges with the aim to study the coherent (0--0.3~\GeVc) and the incoherent photoproduction (0.3--1~\GeVc).
 The choice of the transverse momentum intervals takes into account the broadening of the reconstructed transverse momentum distribution of coherently and incoherently photoproduced \jpsi, mainly due to multiple scattering in the front absorber.
 The raw yield is also extracted in eight \pt intervals up to 8~\GeVc in order to estimate the hadronic contribution as explained below.
 The signal extraction is performed by fitting the invariant mass distribution of the OS dimuons using various combinations of functional forms for the signal and background shapes as discussed in the following.
 The raw \jpsi yield and its statistical uncertainty is then determined as the average of all obtained yield values and corresponding statistical uncertainties, respectively, while the associated systematic uncertainty is taken as the standard deviation of the results.
 The signal is modeled through an extended Crystal Ball function or a pseudo-Gaussian with a mass-dependent width~\cite{ALICE-PUBLIC-2015-006}.
 The non-Gaussian tails were fixed to the values obtained by fitting either a large sample in \pp collisions at \thirteen~\cite{Acharya:2017hjh} or MC simulations where the hadroproduced \jpsi signal is embedded into real events in order to account for detector occupancy effects.
 In the \pt ranges 0--0.3~\GeVc and 0.3--1~\GeVc, additional sets of tails are obtained from MC simulations that use as input coherently and incoherently photoproduced \jpsi from the STARlight MC generator~\cite{Klein:2016yzr}.
 The underlying continuum is described with either a variable-width Gaussian or the ratio of second and third order polynomials~\cite{ALICE:2016flj, ALICE-PUBLIC-2015-006}.

 The nuclear modification factor in the centrality interval $i$ is defined as
 \begin{equation}
     \raa[i](\pt) = \frac{\nraw[i](\pt)}{\br \times \nmb[i] \times \acceff[i,\hh](\pt) \times \taa[i] \times \sigmapp(\pt)},
 \end{equation}
 where \nraw[i] are the measured raw yields, \acceff[i,\hh] is the detector  acceptance and efficiency (assuming unpolarized hadroproduction), \br is the branching ratio to muon pairs~\cite{ParticleDataGroup:2020ssz}, \nmb[i] is the equivalent number of MB events, \taa[i] is the average nuclear overlap function, and \sigmapp is the measured \jpsi cross section in \pp collisions at the same center-of-mass energy~\cite{ALICE:2021qlw}.
 
 The \acceff values are estimated with MC simulations where the \jpsi input \pt and $y$ distributions are adjusted to data, and separately tuned for each centrality class using an iterative procedure.
 The time-dependent status of the electronics channels for the tracking chambers, as well as misalignment of the detector elements, were taken into account.
 The efficiency of the trigger chambers was determined from data and used in the simulations.
 The systematic uncertainty on the \acceff value derives from the uncertainty on the MC input \pt and $y$ distributions and on the tracking, trigger and matching efficiency.
 The former was evaluated by varying the input shapes tuned on data within the statistical uncertainty and by taking into account the correlations between the \pt and $y$ distributions.
 Assuming that the uncertainty related to the correlation does not depend on the collision system and energy, this uncertainty was estimated using a large \pp sample~\cite{Acharya:2017hjh}, by comparing the \acceff values obtained from \pt ($y$) dependent input shapes extracted in narrower $y$ (\pt) intervals with those obtained using the corresponding shapes from the full $y$ and \pt range.
 The remaining uncertainties on the \acceff were determined following the procedure described in detail in Ref.~\cite{Adam:2015isa}.

The normalization to MB events, \nmb[i], is computed as the product of the number of dimuon-triggered events and the inverse of the probability of having a dimuon trigger in a MB event, for the relevant centrality class $i$.
This probability can be obtained with two methods, as explained in Ref.~\cite{ALICE:2016flj}; the difference is taken as the systematic uncertainty.

The average nuclear overlap function \taa and number of participants \Npart (i.e. the number of nucleons in the nuclei undergoing inelastic scattering) are obtained from a Glauber model fit of the \VZERO amplitude~\cite{Adam:2015ptt,dEnterria:2020dwq}. 
The uncertainty on the value of the \VZERO signal amplitude corresponding to the most central 90\% of the total hadronic \PbPb cross section is $\pm 1\%$.
This uncertainty is propagated into the definition of the centrality intervals as explained in Ref.~\cite{ALICE:2016flj}.

The systematic uncertainties on the \raa measurement as a function of centrality are summarized in Table~\ref{tab:syst_central}.

\begin{table}[tb]
\caption{Systematic uncertainties (in percent) on the \raa measurement for different \jpsi \pt intervals. Ranges correspond to the range of values in different centrality classes, whereas the values marked with an asterisk are independent of centrality.}
    \label{tab:syst_central}
    \centering
        \begin{tabular}{|c|c|c|c|}
            \hline
            \pt & 0--0.3~\GeVc & 0.3--1~\GeVc & 1--2~\GeVc \\
            \hline
            Signal extraction & 1.8--3.7 & 1.5--3.4 & 2.4--3.4 \\
            \hline
            MC input & \multicolumn{3}{c|}{2.5} \\ 
            \hline 
            Tracking eff. & \multicolumn{3}{c|}{0--1 + 3*}\\  
            \hline 
            Trigger eff. & 0--1 + 2.8* & 0--1 + 2.0* & 0--1 + 1.5*\\
            \hline
            Matching eff. & \multicolumn{3}{c|}{1*} \\
            \hline
            $N_{\mathrm{MB}}$ & \multicolumn{3}{c|}{0.3*} \\
            \hline
            \taa[] & \multicolumn{3}{c|}{0.7--2.4} \\
            \hline
            Centrality limits & \multicolumn{3}{c|}{0.2--7}\\
            \hline
            \pp cross section & 5.8* & 5.4* & 5.1*\\
            \hline
        \end{tabular}
\end{table}

\begin{figure}[tb]
    \begin{center}
    \includegraphics[width = \textwidth]{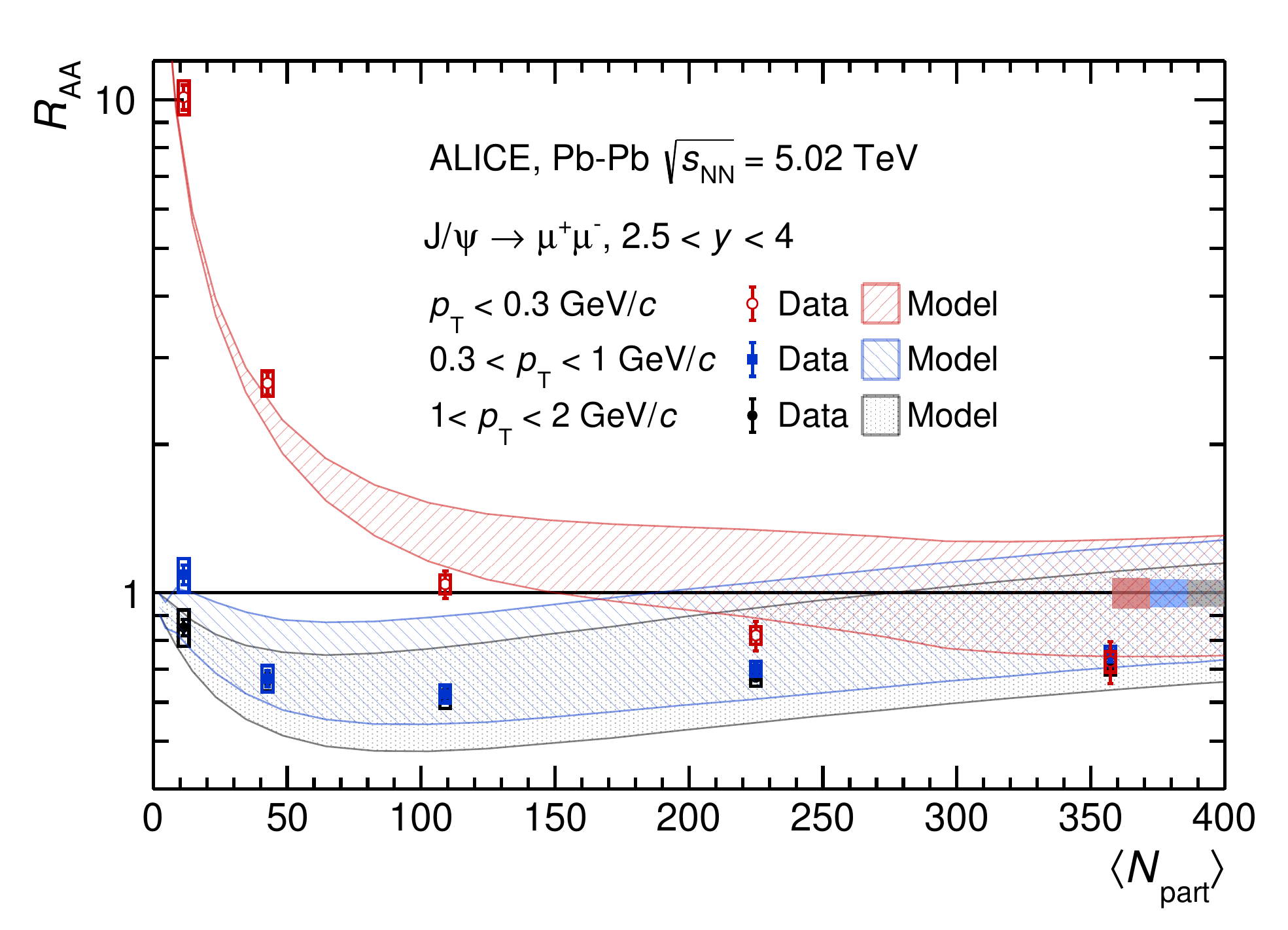}
    \end{center}
    \caption{\jpsi nuclear modification factor as a function of \Npart measured in the rapidity range $2.5 < y < 4$ for three transverse momentum intervals. 
    The vertical bars are the statistical uncertainties and the uncorrelated systematic uncertainties are represented as boxes.
    The centrality-correlated systematic uncertainties are shown as filled boxes at unity.
    Data are compared with predictions from Ref.~\cite{Shi:2017qep}, shown as bands.
    }
    \label{fig:raa}
\end{figure}

Figure~\ref{fig:raa} shows the \raa as a function of the number of participants \Npart.
The relationship between \Npart and centrality is provided in Table~\ref{tab:results}.
The \jpsi \raa for $\pt < 0.3$~\GeVc (where coherent photoproduction would be highest) and $0.3 < \pt < 1.0$~\GeVc (where incoherent photoproduction could contribute) is compared with the \raa for $1.0 < \pt < 2.0$~\GeVc (where hadroproduction dominates).
The \jpsi \raa in the \pt interval 0--0.3~\GeVc is significantly larger than the \raa at larger transverse momenta, except for the most central events.
It reaches a value of about 10 for the most peripheral events.
This large increase is similar to the one of about a factor 7 measured at a lower center-of-mass energy~\cite{Adam:2015gba}.
The measurement in the interval 0.3--1~\GeVc is compatible with the one in 1--2~\GeVc except for the most peripheral events, where it is larger by roughly 2 standard deviations ($\sigma$).
Further studies of the kinematic distribution of this signal could confirm the origin from incoherent photoproduction processes.
Data are compared with a model~\cite{Shi:2017qep} that includes initial \jpsi production, \jpsi regeneration, and a \jpsi photoproduction component for $\pt < 0.3$~\GeVc.
The uncertainty band of the theoretical predictions is mainly due to the variation of the shadowing factor.
QGP effects on the photoproduced \jpsi are taken into account as well.
The theoretical predictions well describe data in the \pt and centrality ranges considered.

The excess with respect to the expected hadronic production was quantified with the same procedure as used in Ref.~\cite{Adam:2015gba}.
For each centrality class, the hadronic \jpsi yield (\dxdpt{N_{\rm AA}^{i,\hh}}) as a function of \pt is parameterized with:
\begin{equation}
 \label{eq:hadronic}
 \dxdpt{N_{\rm AA}^{i,\hh}}(\pt) = \mathcal{N} \times
 \dxdpt{\sigma_{\rm pp}^{\hh}}(\pt) \times \raa[i,\hh](\pt) \times \acceff[i,\hh](\pt).
\end{equation}
The normalization factor $\mathcal{N}$ is defined in such a way that the integral of the function in the \pt interval 1--8~\GeVc is equal to the measured number of \jpsi in the same interval, which is dominated by hadroproduction.
The $\dxdpt{\sigma_{\rm pp}^{\hh}}$ is taken from a fit to the \pp cross section measured by ALICE at \five~\cite{ALICE:2021qlw} with either a power law function~\cite{Bossu:2011qe} or a \LT function~\cite{Tsallis:1987eu,Abelev:2006cs}.
\raa[i,\hh] is a fit to the measured nuclear modification factor as a function of \pt for the same centrality classes as presented above.
For the central to semicentral intervals (0--50\%) a Woods-Saxon like function~\cite{Adam:2015gba} is used, with the parameter $\pt^0$ defining the 50\% crossing point fixed to various values related to the \jpsi mass and average transverse momentum \avpt.
This function was chosen since it can describe the transport model predictions for \jpsi production in heavy-ion collisions~\cite{Zhao:2011cv,Liu:2009nb}.
For the most peripheral intervals (50--90\%), where the recombination effects in the QGP are expected to be smaller, a linear and a constant function are used.
The fit is performed in two \pt intervals 0.65--15~\GeVc and 1--15~\GeVc, where the hadroproduction is the main contribution, and then extrapolated to $\pt=0$.
For the Woods-Saxon function, the quality of the low-\pt extrapolation is assessed by verifying that the functional form reproduces the measured \raa in the most central events where the hadronic contribution is dominant.
Finally, \acceff[i,\hh] is a fit to the hadronic \jpsi acceptance and efficiency for the centrality class $i$, using a ratio of two \LT functions. 
In order not to double-count the uncertainties on the \pp cross section and on the \PbPb \acceff, those were disregarded in the fit to the \raa.
Each combination of different parametrizations and fit ranges results in a different hadronic \jpsi distribution as a function of \pt, which is then integrated in the \pt interval 0--0.3~\GeVc.
The final numbers of expected hadronic \jpsi, defined as the averages of the obtained values, are listed in Table~\ref{tab:results} (fourth column) together with the raw measured numbers of \jpsi (third column).
For the expected hadronic yields, the statistical uncertainty comes from the statistical uncertainty on $\mathcal{N}$, which derives from the statistical uncertainty on the \jpsi raw yield in 1--8~\GeVc.
The systematic uncertainty of the expected yields is taken as the quadratic sum of the standard deviation of the results obtained using different parametrizations and fit ranges, and the average of the individual systematic uncertainties for the variations (including contributions from all factors in Eq.~\ref{eq:hadronic}).

\begin{table}[tb]
\caption{Average number of participants, measured number of \jpsi, estimated number of hadronic \jpsi, difference between these two quantities and resulting \jpsi cross section for coherent photoproduction in the transverse momentum interval 0--0.3~\GeVc for the listed centrality classes. The first quoted uncertainty corresponds to the statistical uncertainty, the second to the centrality uncorrelated systematic uncertainty; in addition, a correlated systematic uncertainty of 6.6\% applies to the cross section in all centrality classes. For the 0--10\% centrality class, the quoted values correspond to a 95\% confidence level interval.}
    \label{tab:results}
    \centering
    \begin{tabular}{|>{\centering\arraybackslash}p{1.6cm}|c|c|c|c|c|}
        \hline
        Centrality class & \Npart & $N_{\mathrm{raw}}^{\jpsi}$ & $N_{\mathrm{hadro}}^{\jpsi}$ & $N_{\mathrm{excess}}^{\jpsi}$ & $\mathrm{d}\sigma_{\mathrm{coh}}^{\jpsi}/\mathrm{d}y~(\mu\mathrm{b})$\\
        \hline
        0--10\% & $357.3 \pm 0.8$ & $8351 \pm 762 \pm 312$ & $8713 \pm 86 \pm 873$ & $<2406$ (95\% CL) & $<230$ (95\% CL) \\
        10--30\% & $225.0 \pm 1.2$ & $9624 \pm 571 \pm 278$ & $8274 \pm 60 \pm 742$ & $1350 \pm 574 \pm 792$ & $145 \pm 62 \pm 85$ \\
        30--50\% & $109.0 \pm 1.1$ & $4280 \pm 225 \pm 105$ & $2562 \pm 23 \pm 178$ & $1718 \pm 226 \pm 207$ & $179 \pm 24 \pm 22$ \\
        50--70\% & $42.7 \mp 0.7$ & $2763 \pm 98 \pm 68$ & $674 \pm 8 \pm 40$ & $2089 \pm 98 \pm 79$ & $216 \pm 10 \pm 12 $ \\
        70--90\% & $11.3 \pm 0.2$ & $1758 \pm 57 \pm 32$ & $138 \pm 3 \pm 9$ & $1620 \pm 57 \pm 33$ & $167 \pm 6 \pm 12$ \\
        \hline
    \end{tabular}
\end{table}

The estimated number of hadroproduced \jpsi is subtracted from the measured raw signal to obtain the number of \jpsi in excess (fifth column of Table~\ref{tab:results}).
The measured number of \jpsi exceeds the hadronic production by 24$\sigma$ in the 70--90\% centrality class, 16$\sigma$ in  50--70\%, 5.6$\sigma$ in 30--50\% and 1.4$\sigma$ in 10--30\%.
A 95\% confidence interval when combining all uncertainties is provided in the centrality class 0--10\% where no significant excess is observed within the current experimental uncertainties.

Assuming that the underlying process for the \jpsi excess is photoproduction, the number of coherently photoproduced \jpsi in $0<\pt<0.3$~\GeVc can be extracted after correcting the excess yield for the fractions of \jpsi from incoherent photoproduction (\ffi) and from the decay of coherently photoproduced $\psi$(2S) (\ffd) as described in Ref.~\cite{ALICE:2019tqa}.
Those fractions were measured in UPC collisions at the same center-of-mass energy, although in a slightly different \pt interval, $\pt<0.25$~\GeVc~\cite{ALICE:2019tqa}.
They were therefore recomputed for $\pt<0.3$~\GeVc.
The corresponding values and systematic uncertainties are $\ffi = 0.089 \pm 0.034$ and $\ffd = 0.066 \pm 0.013$.
In the following it was assumed that these fractions are the same in UPC and hadronic collisions and that they do not depend significantly on the collision centrality.
The first assumption seems realistic for \ffd, although \ffi might vary if the coherence is incomplete in the presence of hadronic interactions.

Finally, the cross section is obtained by correcting the excess yield for the branching ratio to OS dimuons, for the \acceff factor estimated by means of STARlight~\cite{Klein:2016yzr} simulations embedded into data for each centrality class, taking into account that the coherently photoproduced \jpsi mesons are expected to be transversely polarized, and by normalizing to the integrated luminosity and the width of the rapidity range.
The systematic uncertainties are summarized in Table~\ref{tab:syst_xsec}.
The uncertainties on the number of excess \jpsi are discussed above. 
The contributions from the \acceff are the same as in Table~\ref{tab:syst_central}, except for the one on the STARlight MC input, which is obtained as described in Ref.~\cite{ALICE:2019tqa}.
An additional systematic uncertainty of 2\% due to the transverse momentum resolution was estimated by comparing the \acceff obtained with or without the \pt selection at 0.3~\GeVc.
The systematic uncertainty on the luminosity mainly originates from the uncertainty of the reference \VZERO trigger cross section measured with van der Meer scans~\cite{ALICE-PUBLIC-2021-001}.
The uncertainties on \ffi and \ffd are estimated as described in Ref.~\cite{ALICE:2019tqa}.

\begin{table}[tb]
\caption{Systematic uncertainties on the coherent \jpsi cross section (notation is the same as in Table~\ref{tab:syst_central}).}
    \label{tab:syst_xsec}
    \centering
    \begin{tabular}{|c|c|}
            \hline
            Source & Value (\%)\\
            \hline
            Branching Ratio & 0.5* \\
            \hline
            $N_{\jpsi}^{\rm excess}$ & 2--58.7   \\
            \hline
            \ffi & 2.9* \\
            \hline
            \ffd & 1.1* \\
            \hline
            Tracking eff. & 0--0.5 + 3* \\  
            \hline 
            Trigger eff. & 0--0.5 + 3.6* \\
            \hline
            Matching eff. & 1* \\
            \hline
            MC input & 0.1* \\
            \hline
            \pt selection & 2* \\
            \hline
            Centrality limits &  0.2--7\\
            \hline
            $\mathcal{L}_{\mathrm{int}}$ & 2.5*\\
            \hline
    \end{tabular}
\end{table}

\begin{figure}[tb]
    \centering
    \includegraphics[width=0.9\textwidth]{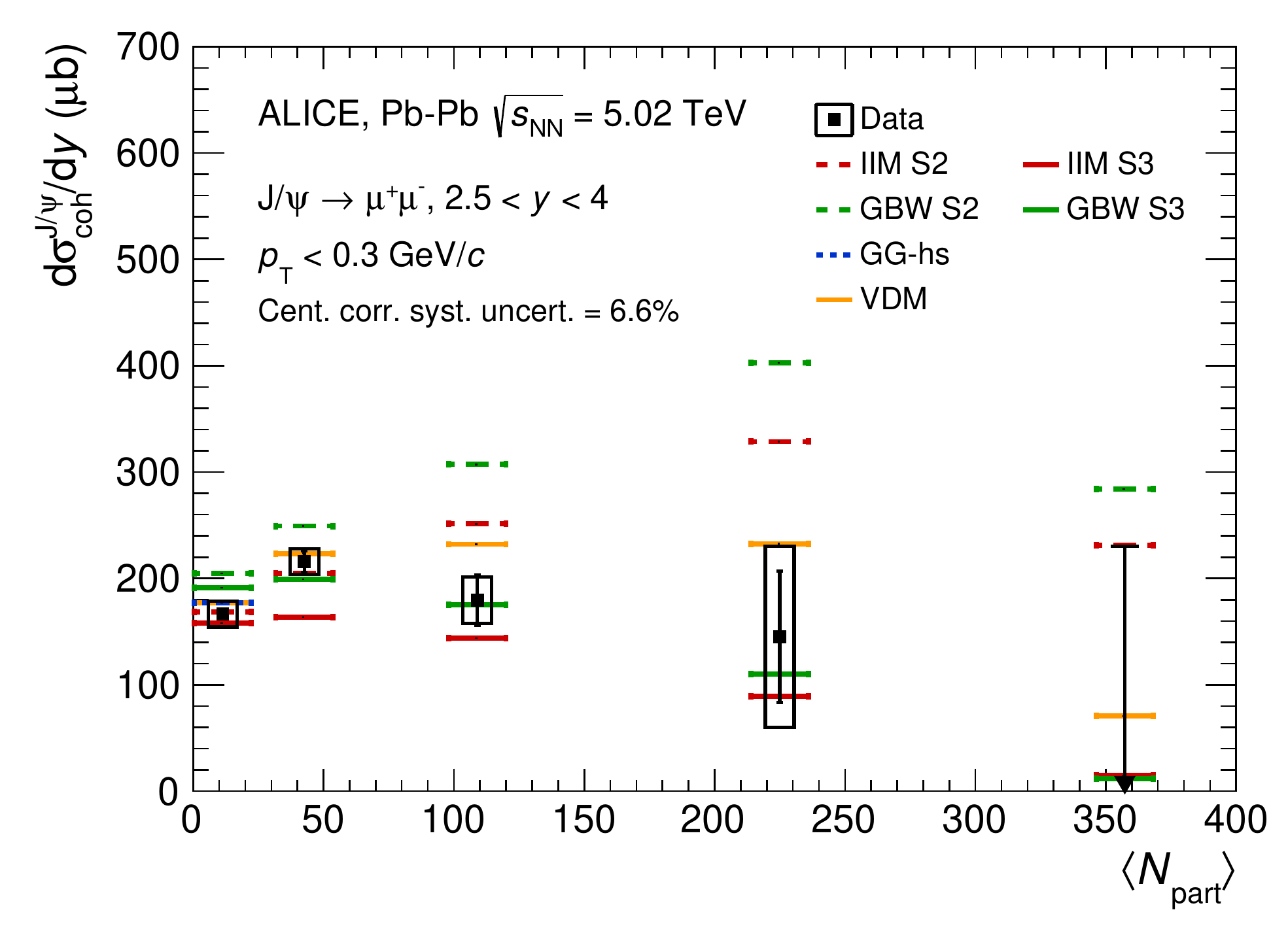}
    \caption{\jpsi coherent photoproduction cross section as a function of \Npart at forward rapidity in \PbPb collisions at \fivenn. 
    The vertical bars are the statistical uncertainties and the uncorrelated systematic uncertainties are represented as boxes.
    The centrality correlated systematic uncertainties are quoted in the legend. Results are compared with theoretical calculations from Ref.~\cite{Cepila:2017nef} (GG-hs), Ref.~\cite{GayDucati:2018who} (IIM S2 and S3, and GBW S2 and S3) and from Ref.~\cite{Klusek-Gawenda:2015hja} with updated Glauber calculations from Ref.~\cite{Klusek-Gawenda:2018zfz} (VDM).
    The figure shows the integral of the cross section measurement as well as the corresponding theoretical model values in each centrality interval. Note that the most central bin, where only an upper limit is given, is half the size of the other intervals. Therefore, to evaluate the centrality dependence of \jpsi coherent photoproduction, both data and theory have to be multiplied by a factor of two.}
    \label{fig:cross_section}
\end{figure}

The coherent \jpsi photoproduction cross section at \fivenn as a function of \Npart is shown in Figure~\ref{fig:cross_section}.
Empty boxes correspond to the uncorrelated systematic uncertainties.
The correlated systematic uncertainty amounts to 6.6\%, independent of centrality, and is quoted in the legend.

The result is compared with theoretical calculations that use an effective description based on UPC color dipole models.
The GG-hs calculations~\cite{Cepila:2017nef} are based on models representing subnucleonic degrees of freedom as hot spots, whose number increases with increasing photon-target center of mass energy. The calculation is extended from protonic to nuclear targets using Glauber--Gribov formalism (GG)~\cite{Cepila:2017nef}.
The photon flux is estimated in the same way as in the UPC case, but the integral is limited to the impact parameter range of the selected centrality class.
The calculation from Ref.~\cite{Klusek-Gawenda:2018zfz} is based on a vector dominance model, in which the photon fluctuates into a vector meson component that propagates through the nucleus and fragments into an on-shell vector meson.
In this model, which will be referred to as VDM in the following, the photon flux is modified with respect to the one used in UPC calculations by considering only the photons that reach the geometrical region of the target nucleus outside of the overlap region.
In the GBW calculation, the light cone color dipole formalism is used, while the IIM calculation is based on the Color Glass Condensate approach~\cite{GayDucati:2018who}.
The GBW and IIM calculations provide two scenarios.
In the first one (called S2 in Ref.~\cite{GayDucati:2018who}), the photon flux is modified in a similar way as for the VDM model.
However, in contrast with the latter, an effective area is used in building the flux, which disregards the region of nuclear overlap.
This prevents the flux and the resulting cross section from being progressively reduced towards more central collisions.
In the second scenario (S3), an additional modification of the photon-nucleus cross section is introduced, in which the overlap region between the two nuclei is assumed not to contribute to coherent photoproduction resulting in significant reduction of the photoproduction cross sections towards more central collisions.

The hot-spot model prediction (GG-hs) is only available for the most peripheral centrality interval (70-90\%) where the calculation is compatible with data.
The other models provide predictions for all centrality intervals.
The VDM model predicts a mild increase of the cross section in peripheral events, a flat evolution in semi-central events, and a decrease of the cross section in the most central events, in fair agreement with data.
Notice that the figure shows the integral of the cross section in each centrality interval and the most central interval is half the size of the others.
If one accounts for the interval width, the predictions for the most central interval would be twice as large, resulting in a rather mild decrease of the cross section with centrality.
This model uses an optical Glauber model to describe the collision centrality, but a similar agreement with data can be obtained with a simplified relation between impact parameter and centrality~\cite{Klusek-Gawenda:2015hja}.
The IIM and GBW models with unmodified photon-nucleus cross section (S2) predict a steady increase of the \jpsi coherent photoproduction cross section with centrality, once the width of the centrality intervals is properly accounted for.
In data, this increasing trend is only observed for the two most peripheral intervals.
In this scenario, the GBW model overestimates the data in all centrality intervals.
The IIM model is in agreement with data in the first two centrality intervals, while it starts to deviate from the data by 2.1$\sigma$ in the 30--50\% centrality interval.
The S3 version of the GBW and IIM models~\cite{GayDucati:2018who} excluding the nuclear overlap region from the photon-nucleus cross section calculation predicts a decrease of the cross section from semicentral to central events (similar to the one of Ref.~\cite{Klusek-Gawenda:2018zfz}, which, however, requires only a modification of the photon flux), and is compatible with the data in the full centrality range considering the current uncertainties.
Since the transverse momentum of the coherently photoproduced vector meson is of the order of the inverse of the target size, the interaction occurring with the remaining nucleus fragment outside the overlap area would result in a larger average \pt and a wider \pt distribution for the photoproduced \jpsi.
A measurement of the \jpsi \pt distribution at low \pt is therefore needed to clarify what is the underlying mechanism leading to the observed distribution as a function of centrality.

\begin{figure}[tb]
    \begin{center}
    \includegraphics[width = \textwidth]{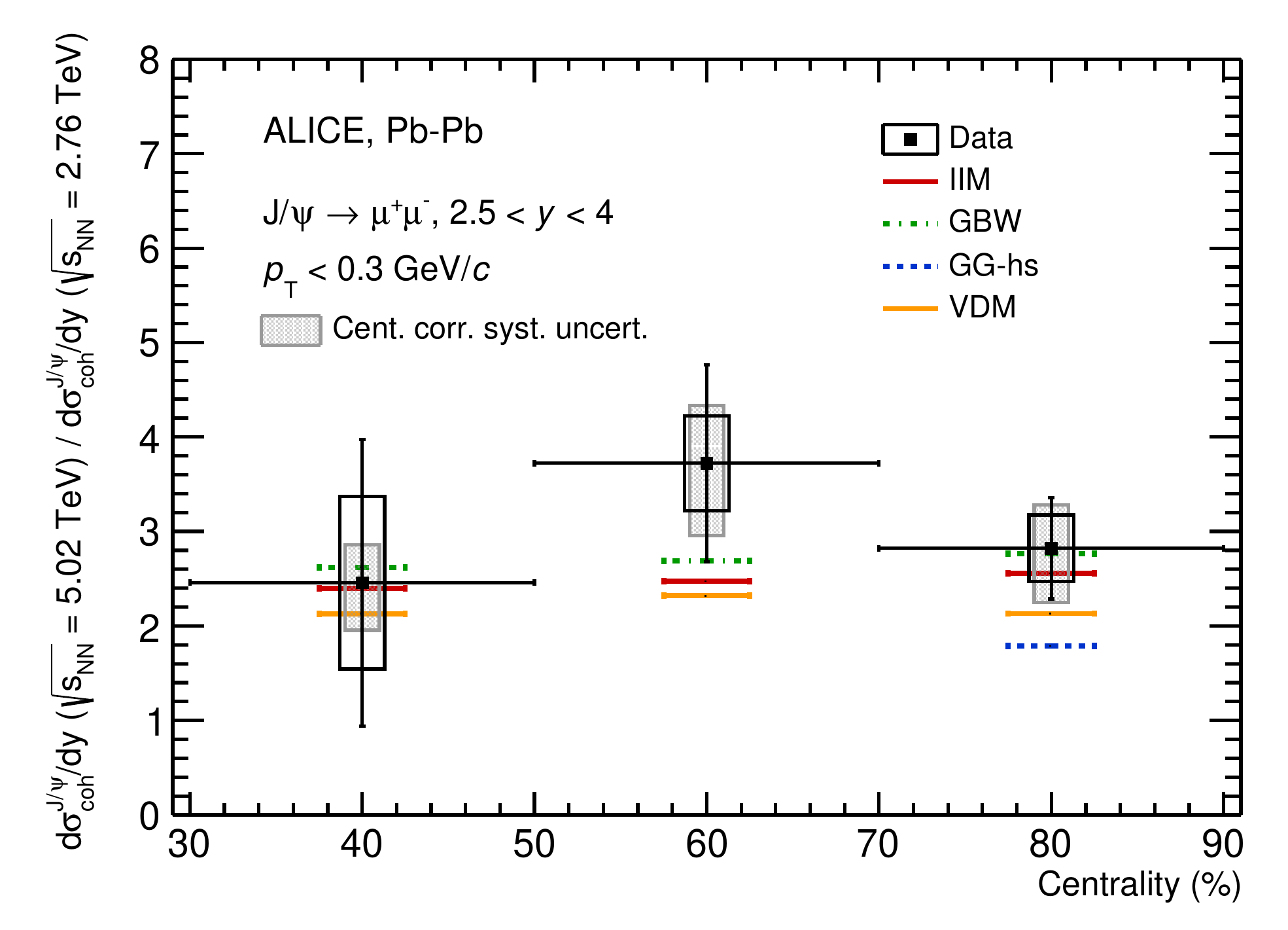}
    \end{center}
   \caption{\jpsi coherent photoproduction cross section ratio for two different energies (\fivenn over \twosevensixnn) as a function of centrality. The data at \twosevensixnn are taken from Ref.~\cite{Adam:2015gba}. The vertical lines are the statistical uncertainties while the open (filled) boxes are the centrality uncorrelated (correlated) systematic uncertainties.
   Results are compared with theoretical calculations from Ref.~\cite{Cepila:2017nef} (GG-hs), Ref.~\cite{GayDucati:2018who} (IIM and GBW) and from Ref.~\cite{Klusek-Gawenda:2015hja} with updated Glauber calculations from Ref.~\cite{Klusek-Gawenda:2018zfz} (VDM).}
    \label{fig:cross_section_ratio}
\end{figure}

The models described here provided predictions also for the measurement at \twosevensixnn~\cite{Adam:2015gba}.
The corresponding figure can be found in the Appendix~\ref{sec:appendix}.
The ratio of the measurements at \fivenn and \twosevensixnn~\cite{Adam:2015gba} is shown in Fig.~\ref{fig:cross_section_ratio}.
In the ratio, only the systematic uncertainty on the branching ratio cancels out.
The centrality uncorrelated (correlated) systematic uncertainties in Table~\ref{tab:syst_xsec} are represented as open (filled) boxes in Fig.~\ref{fig:cross_section_ratio}.
The centrality correlated uncertainties are mainly due to the uncertainty on \ffi and \ffd, which were asymmetric in the estimation performed at \twosevensixnn.
The cross section increase with the center-of-mass energy does not depend significantly on the centrality.
Figure~\ref{fig:cross_section_ratio} shows that the hot-spot model tends to underpredict the increase of the cross section with the center-of-mass energy in peripheral hadronic interactions, while the other models are in fair agreement with the measured ratio in all centrality ranges within the large uncertainties.
For the IIM and GBW models no distinction is done in this case between the scenarios with or without modification of the photon-nucleus cross section since their energy dependence is exactly the same.

In summary, this Letter reports the measurement of \jpsi production at very low \pt as a function of centrality in hadronic \PbPb collisions at \fivenn at forward rapidity.
The nuclear modification factor \raa shows a large enhancement of the \jpsi yield for $\pt < 0.3$~\GeVc with respect to expectations from hadronic production.
This excess, which was previously seen in more peripheral collisions, is now confirmed to be present for most of the total hadronic cross section, including in collisions with a large nuclear overlap, down to at least a level of 30\% in centrality.
The enhancement has a significance of 24$\sigma$ in the 70--90\% centrality class, 16$\sigma$ in 50--70\% and 5.6$\sigma$ in the centrality class 30--50\%.
The reported observation extends previous measurements performed by the ALICE, LHCb and STAR Collaborations, supporting coherent photoproduction in hadronic collisions as the underlying mechanism.
Based on this assumption, the corresponding cross section is extracted for the centrality classes 10--30\%, 30--50\%, 50--70\% and 70--90\% while an upper limit is given for 0--10\%.
The ratio of coherent photoproduction cross sections for \fivenn over \twosevensixnn is extracted as a function of centrality and shows a flat dependence on centrality within uncertainties.
A set of theoretical calculations successfully used to describe coherent photoproduction in UPC, and modified to account for geometrical constraints on the photon flux in the selected centrality classes, is compared with the measurement.
The cross section as a function of centrality is well described by two models, one implementing a modification of the photon flux only~\cite{Klusek-Gawenda:2018zfz}, and the other requiring an additional modification of the photon-nucleus cross section~\cite{GayDucati:2018who}.
Additional measurements of the \pt-differential photoproduction cross section as a function of centrality and further comparison with models using different photoproduction scenarios would help to clarify the effect of the disruption of the nucleus and nucleons by hadronic interactions on the coherence condition of vector meson photoproduction.


\newenvironment{acknowledgement}{\relax}{\relax}
\begin{acknowledgement}
\section*{Acknowledgements}

The ALICE Collaboration would like to thank all its engineers and technicians for their invaluable contributions to the construction of the experiment and the CERN accelerator teams for the outstanding performance of the LHC complex.
The ALICE Collaboration gratefully acknowledges the resources and support provided by all Grid centres and the Worldwide LHC Computing Grid (WLCG) collaboration.
The ALICE Collaboration acknowledges the following funding agencies for their support in building and running the ALICE detector:
A. I. Alikhanyan National Science Laboratory (Yerevan Physics Institute) Foundation (ANSL), State Committee of Science and World Federation of Scientists (WFS), Armenia;
Austrian Academy of Sciences, Austrian Science Fund (FWF): [M 2467-N36] and Nationalstiftung f\"{u}r Forschung, Technologie und Entwicklung, Austria;
Ministry of Communications and High Technologies, National Nuclear Research Center, Azerbaijan;
Conselho Nacional de Desenvolvimento Cient\'{\i}fico e Tecnol\'{o}gico (CNPq), Financiadora de Estudos e Projetos (Finep), Funda\c{c}\~{a}o de Amparo \`{a} Pesquisa do Estado de S\~{a}o Paulo (FAPESP) and Universidade Federal do Rio Grande do Sul (UFRGS), Brazil;
Bulgarian Ministry of Education and Science, within the National Roadmap for Research Infrastructures 2020-2027 (object CERN), Bulgaria;
Ministry of Education of China (MOEC) , Ministry of Science \& Technology of China (MSTC) and National Natural Science Foundation of China (NSFC), China;
Ministry of Science and Education and Croatian Science Foundation, Croatia;
Centro de Aplicaciones Tecnol\'{o}gicas y Desarrollo Nuclear (CEADEN), Cubaenerg\'{\i}a, Cuba;
Ministry of Education, Youth and Sports of the Czech Republic, Czech Republic;
The Danish Council for Independent Research | Natural Sciences, the VILLUM FONDEN and Danish National Research Foundation (DNRF), Denmark;
Helsinki Institute of Physics (HIP), Finland;
Commissariat \`{a} l'Energie Atomique (CEA) and Institut National de Physique Nucl\'{e}aire et de Physique des Particules (IN2P3) and Centre National de la Recherche Scientifique (CNRS), France;
Bundesministerium f\"{u}r Bildung und Forschung (BMBF) and GSI Helmholtzzentrum f\"{u}r Schwerionenforschung GmbH, Germany;
General Secretariat for Research and Technology, Ministry of Education, Research and Religions, Greece;
National Research, Development and Innovation Office, Hungary;
Department of Atomic Energy Government of India (DAE), Department of Science and Technology, Government of India (DST), University Grants Commission, Government of India (UGC) and Council of Scientific and Industrial Research (CSIR), India;
National Research and Innovation Agency - BRIN, Indonesia;
Istituto Nazionale di Fisica Nucleare (INFN), Italy;
Japanese Ministry of Education, Culture, Sports, Science and Technology (MEXT) and Japan Society for the Promotion of Science (JSPS) KAKENHI, Japan;
Consejo Nacional de Ciencia (CONACYT) y Tecnolog\'{i}a, through Fondo de Cooperaci\'{o}n Internacional en Ciencia y Tecnolog\'{i}a (FONCICYT) and Direcci\'{o}n General de Asuntos del Personal Academico (DGAPA), Mexico;
Nederlandse Organisatie voor Wetenschappelijk Onderzoek (NWO), Netherlands;
The Research Council of Norway, Norway;
Commission on Science and Technology for Sustainable Development in the South (COMSATS), Pakistan;
Pontificia Universidad Cat\'{o}lica del Per\'{u}, Peru;
Ministry of Education and Science, National Science Centre and WUT ID-UB, Poland;
Korea Institute of Science and Technology Information and National Research Foundation of Korea (NRF), Republic of Korea;
Ministry of Education and Scientific Research, Institute of Atomic Physics, Ministry of Research and Innovation and Institute of Atomic Physics and University Politehnica of Bucharest, Romania;
Ministry of Education, Science, Research and Sport of the Slovak Republic, Slovakia;
National Research Foundation of South Africa, South Africa;
Swedish Research Council (VR) and Knut \& Alice Wallenberg Foundation (KAW), Sweden;
European Organization for Nuclear Research, Switzerland;
Suranaree University of Technology (SUT), National Science and Technology Development Agency (NSTDA) and National Science, Research and Innovation Fund (NSRF via PMU-B B05F650021), Thailand;
Turkish Energy, Nuclear and Mineral Research Agency (TENMAK), Turkey;
National Academy of  Sciences of Ukraine, Ukraine;
Science and Technology Facilities Council (STFC), United Kingdom;
National Science Foundation of the United States of America (NSF) and United States Department of Energy, Office of Nuclear Physics (DOE NP), United States of America.
In addition, individual groups or members have received support from:
Marie Sk\l{}odowska Curie, Strong 2020 - Horizon 2020, European Research Council (grant nos. 824093, 896850, 950692), European Union;
Academy of Finland (Center of Excellence in Quark Matter) (grant nos. 346327, 346328), Finland;
Programa de Apoyos para la Superaci\'{o}n del Personal Acad\'{e}mico, UNAM, Mexico;
\end{acknowledgement}

\bibliographystyle{utphys}   
\bibliography{bibliography}

\newpage
\appendix
\section{\texorpdfstring{\jpsi}{J/psi} photoproduction in \PbPb collisions at \texorpdfstring{\twosevensixnn}{sqrt(s\_NN = 2.76 TeV}}
\label{sec:appendix}

\begin{figure}[tb]
    \centering
    \includegraphics[width = 0.9\textwidth]{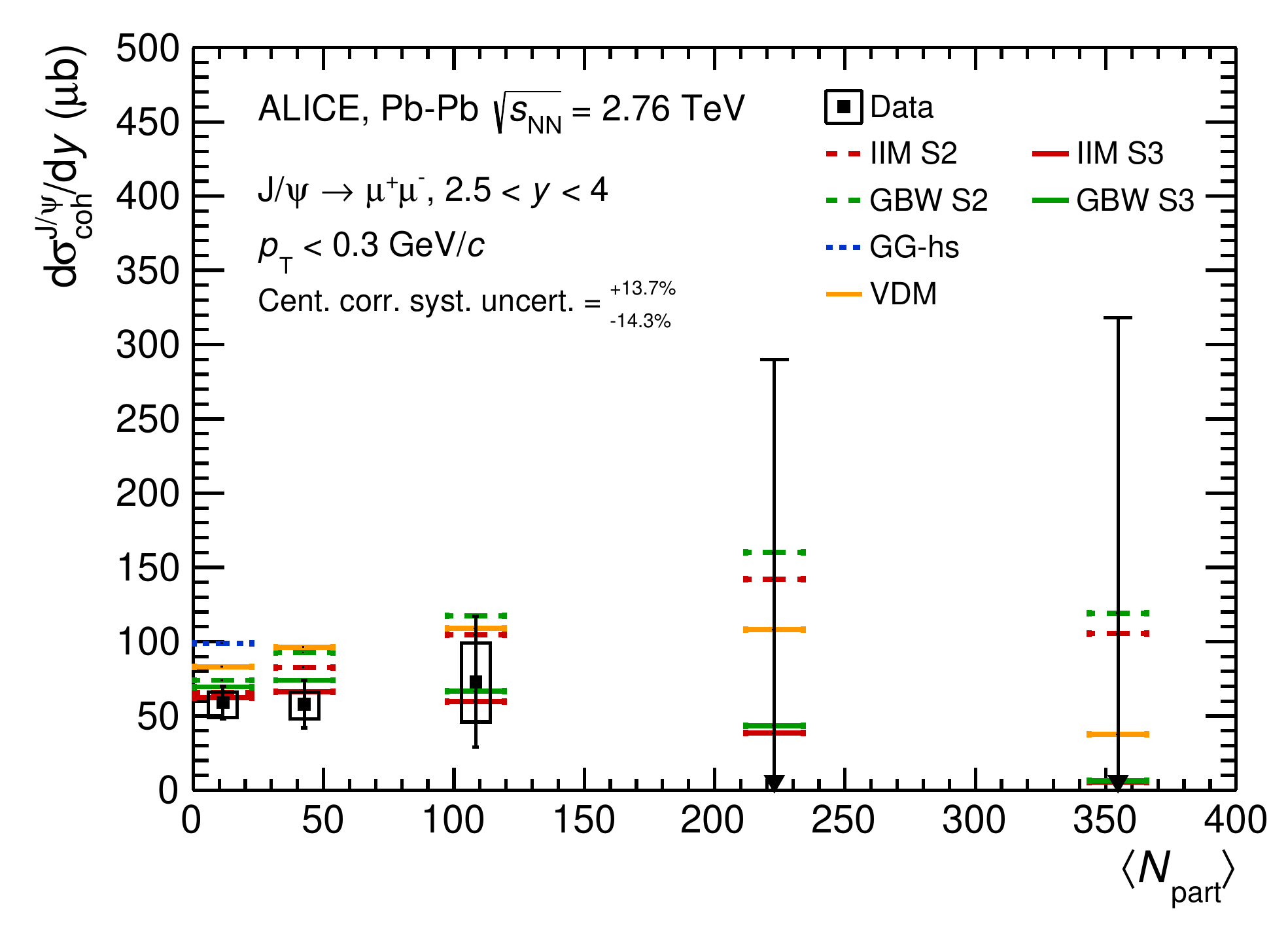}
    \caption{\jpsi coherent photoproduction cross section as a function of \Npart at forward rapidity in \PbPb collisions at \twosevensixnn~\cite{Adam:2015gba}.
    The vertical bars are the statistical uncertainties and the uncorrelated systematic uncertainties are represented as boxes.
    The centrality correlated systematic uncertainties are quoted in the legend. 
    Results are compared with theoretical calculations from Ref.~\cite{Cepila:2017nef} (GG-hs), Ref.~\cite{GayDucati:2018who} (IIM S2 and S3, and GBW S2 and S3) and from Ref.~\cite{Klusek-Gawenda:2015hja} with updated Glauber calculations from Ref.~\cite{Klusek-Gawenda:2018zfz} (VDM).
    The figure shows the integral of the cross section measurement as well as the corresponding theoretical model values in each centrality interval. Note that the most central bin, where only an upper limit is given, is half the size of the other intervals. Therefore, to evaluate the centrality dependence of \jpsi coherent photoproduction, both data and theory have to be multiplied by a factor of two.
    }
    \label{fig:cross_section_276}
\end{figure}

Figure~\ref{fig:cross_section_276} shows the coherent photoproduction measured at \twosevensixnn~\cite{Adam:2015gba}.
Empty boxes correspond to the uncorrelated systematic uncertainties.
The centrality correlated systematic uncertainty mainly comes from the uncertainties on \ffi and \ffd, which are asymmetric.

The data are compared with predictions from the same set of models that were described in detail in the paper to which this appendix is associated.
The hot-spot model prediction (GG-hs)~\cite{Cepila:2017nef} is only available for the most peripheral centrality interval (70-90\%) and it is found to overestimate the data.
The other predictions are available for all centrality intervals.
The centrality dependence of the models is similar to the one shown at \fivenn.
The IIM and GBW predictions~\cite{GayDucati:2018who} steadily increase with centrality in the scenario with unmodified photon-nucleus cross section (S2), while the use of an effective cross section where the overlap region between the two nuclei is assumed not to contribute to coherent photoproduction (S3) results in a reduction of the cross section toward more central collisions.
However, both scenarios are compatible with data in the current uncertainties.
Finally, the VDM calculations are in  agreement with data in the most central events while they tend to overestimate data in the 50--70\% and 70--90\% centrality bins.

%
%
\section{The ALICE Collaboration}
\label{app:collab}
\begin{flushleft} 
\small

S.~Acharya\,\orcidlink{0000-0002-9213-5329}\,$^{\rm 123,131}$, 
D.~Adamov\'{a}\,\orcidlink{0000-0002-0504-7428}\,$^{\rm 85}$, 
A.~Adler$^{\rm 69}$, 
G.~Aglieri Rinella\,\orcidlink{0000-0002-9611-3696}\,$^{\rm 32}$, 
M.~Agnello\,\orcidlink{0000-0002-0760-5075}\,$^{\rm 29}$, 
N.~Agrawal\,\orcidlink{0000-0003-0348-9836}\,$^{\rm 50}$, 
Z.~Ahammed\,\orcidlink{0000-0001-5241-7412}\,$^{\rm 131}$, 
S.~Ahmad\,\orcidlink{0000-0003-0497-5705}\,$^{\rm 15}$, 
S.U.~Ahn\,\orcidlink{0000-0001-8847-489X}\,$^{\rm 70}$, 
I.~Ahuja\,\orcidlink{0000-0002-4417-1392}\,$^{\rm 37}$
% ...

\section*{Affiliation Notes}

$^{\rm I}$ Deceased\\
$^{\rm II}$ Also at: Max-Planck-Institut f\"{u}r Physik, Munich, Germany\\
$^{\rm III}$ Also at: Italian National Agency for New Technologies, Energy and Sustainable Economic Development (ENEA), Bologna, Italy\\
$^{\rm IV}$ Also at: Dipartimento DET del Politecnico di Torino, Turin, Italy\\
$^{\rm V}$ Also at: Department of Applied Physics, Aligarh Muslim University, Aligarh, India\\
$^{\rm VI}$ Also at: Institute of Theoretical Physics, University of Wroclaw, Poland\\
$^{\rm VII}$ Also at: An institution covered by a cooperation agreement with CERN\\

\section*{Collaboration Institutes}

$^{1}$ A.I. Alikhanyan National Science Laboratory (Yerevan Physics Institute) Foundation, Yerevan, Armenia\\
$^{2}$ AGH University of Krakow, Cracow, Poland\\
$^{3}$ Bogolyubov Institute for Theoretical Physics, National Academy of Sciences of Ukraine, Kiev, Ukraine\\
$^{4}$ Bose Institute, Department of Physics  and Centre for Astroparticle Physics and Space Science (CAPSS), Kolkata, India\\
$^{5}$ California Polytechnic State University, San Luis Obispo, California, United States\\
$^{6}$ Central China Normal University, Wuhan, China\\
$^{7}$ Centro de Aplicaciones Tecnol\'{o}gicas y Desarrollo Nuclear (CEADEN), Havana, Cuba\\
$^{8}$ Centro de Investigaci\'{o}n y de Estudios Avanzados (CINVESTAV), Mexico City and M\'{e}rida, Mexico\\
$^{9}$ Chicago State University, Chicago, Illinois, United States\\
$^{10}$ China Institute of Atomic Energy, Beijing, China\\
$^{11}$ Chungbuk National University, Cheongju, Republic of Korea\\
$^{12}$ Comenius University Bratislava, Faculty of Mathematics, Physics and Informatics, Bratislava, Slovak Republic\\
$^{13}$ COMSATS University Islamabad, Islamabad, Pakistan\\
$^{14}$ Creighton University, Omaha, Nebraska, United States\\
$^{15}$ Department of Physics, Aligarh Muslim University, Aligarh, India\\
$^{16}$ Department of Physics, Pusan National University, Pusan, Republic of Korea\\
$^{17}$ Department of Physics, Sejong University, Seoul, Republic of Korea\\
$^{18}$ Department of Physics, University of California, Berkeley, California, United States\\
$^{19}$ Department of Physics, University of Oslo, Oslo, Norway\\
$^{20}$ Department of Physics and Technology, University of Bergen, Bergen, Norway\\
$^{21}$ Dipartimento di Fisica, Universit\`{a} di Pavia, Pavia, Italy\\
$^{22}$ Dipartimento di Fisica dell'Universit\`{a} and Sezione INFN, Cagliari, Italy\\
$^{23}$ Dipartimento di Fisica dell'Universit\`{a} and Sezione INFN, Trieste, Italy\\
$^{24}$ Dipartimento di Fisica dell'Universit\`{a} and Sezione INFN, Turin, Italy\\
$^{25}$ Dipartimento di Fisica e Astronomia dell'Universit\`{a} and Sezione INFN, Bologna, Italy\\
$^{26}$ Dipartimento di Fisica e Astronomia dell'Universit\`{a} and Sezione INFN, Catania, Italy\\
$^{27}$ Dipartimento di Fisica e Astronomia dell'Universit\`{a} and Sezione INFN, Padova, Italy\\
$^{28}$ Dipartimento di Fisica `E.R.~Caianiello' dell'Universit\`{a} and Gruppo Collegato INFN, Salerno, Italy\\
$^{29}$ Dipartimento DISAT del Politecnico and Sezione INFN, Turin, Italy\\
$^{30}$ Dipartimento di Scienze MIFT, Universit\`{a} di Messina, Messina, Italy\\
$^{31}$ Dipartimento Interateneo di Fisica `M.~Merlin' and Sezione INFN, Bari, Italy\\
$^{32}$ European Organization for Nuclear Research (CERN), Geneva, Switzerland\\
$^{33}$ Faculty of Electrical Engineering, Mechanical Engineering and Naval Architecture, University of Split, Split, Croatia\\
$^{34}$ Faculty of Engineering and Science, Western Norway University of Applied Sciences, Bergen, Norway\\
$^{35}$ Faculty of Nuclear Sciences and Physical Engineering, Czech Technical University in Prague, Prague, Czech Republic\\
$^{36}$ Faculty of Physics, Sofia University, Sofia, Bulgaria\\
$^{37}$ Faculty of Science, P.J.~\v{S}af\'{a}rik University, Ko\v{s}ice, Slovak Republic\\
$^{38}$ Frankfurt Institute for Advanced Studies, Johann Wolfgang Goethe-Universit\"{a}t Frankfurt, Frankfurt, Germany\\
$^{39}$ Fudan University, Shanghai, China\\
$^{40}$ Gangneung-Wonju National University, Gangneung, Republic of Korea\\
$^{41}$ Gauhati University, Department of Physics, Guwahati, India\\
$^{42}$ Helmholtz-Institut f\"{u}r Strahlen- und Kernphysik, Rheinische Friedrich-Wilhelms-Universit\"{a}t Bonn, Bonn, Germany\\
$^{43}$ Helsinki Institute of Physics (HIP), Helsinki, Finland\\
$^{44}$ High Energy Physics Group,  Universidad Aut\'{o}noma de Puebla, Puebla, Mexico\\
$^{45}$ Horia Hulubei National Institute of Physics and Nuclear Engineering, Bucharest, Romania\\
$^{46}$ Indian Institute of Technology Bombay (IIT), Mumbai, India\\
$^{47}$ Indian Institute of Technology Indore, Indore, India\\
$^{48}$ INFN, Laboratori Nazionali di Frascati, Frascati, Italy\\
$^{49}$ INFN, Sezione di Bari, Bari, Italy\\
$^{50}$ INFN, Sezione di Bologna, Bologna, Italy\\
$^{51}$ INFN, Sezione di Cagliari, Cagliari, Italy\\
$^{52}$ INFN, Sezione di Catania, Catania, Italy\\
$^{53}$ INFN, Sezione di Padova, Padova, Italy\\
$^{54}$ INFN, Sezione di Pavia, Pavia, Italy\\
$^{55}$ INFN, Sezione di Torino, Turin, Italy\\
$^{56}$ INFN, Sezione di Trieste, Trieste, Italy\\
$^{57}$ Inha University, Incheon, Republic of Korea\\
$^{58}$ Institute for Gravitational and Subatomic Physics (GRASP), Utrecht University/Nikhef, Utrecht, Netherlands\\
$^{59}$ Institute of Experimental Physics, Slovak Academy of Sciences, Ko\v{s}ice, Slovak Republic\\
$^{60}$ Institute of Physics, Homi Bhabha National Institute, Bhubaneswar, India\\
$^{61}$ Institute of Physics of the Czech Academy of Sciences, Prague, Czech Republic\\
$^{62}$ Institute of Space Science (ISS), Bucharest, Romania\\
$^{63}$ Institut f\"{u}r Kernphysik, Johann Wolfgang Goethe-Universit\"{a}t Frankfurt, Frankfurt, Germany\\
$^{64}$ Instituto de Ciencias Nucleares, Universidad Nacional Aut\'{o}noma de M\'{e}xico, Mexico City, Mexico\\
$^{65}$ Instituto de F\'{i}sica, Universidade Federal do Rio Grande do Sul (UFRGS), Porto Alegre, Brazil\\
$^{66}$ Instituto de F\'{\i}sica, Universidad Nacional Aut\'{o}noma de M\'{e}xico, Mexico City, Mexico\\
$^{67}$ iThemba LABS, National Research Foundation, Somerset West, South Africa\\
$^{68}$ Jeonbuk National University, Jeonju, Republic of Korea\\
$^{69}$ Johann-Wolfgang-Goethe Universit\"{a}t Frankfurt Institut f\"{u}r Informatik, Fachbereich Informatik und Mathematik, Frankfurt, Germany\\
$^{70}$ Korea Institute of Science and Technology Information, Daejeon, Republic of Korea\\
$^{71}$ KTO Karatay University, Konya, Turkey\\
$^{72}$ Laboratoire de Physique Subatomique et de Cosmologie, Universit\'{e} Grenoble-Alpes, CNRS-IN2P3, Grenoble, France\\
$^{73}$ Lawrence Berkeley National Laboratory, Berkeley, California, United States\\
$^{74}$ Lund University Department of Physics, Division of Particle Physics, Lund, Sweden\\
$^{75}$ Nagasaki Institute of Applied Science, Nagasaki, Japan\\
$^{76}$ Nara Women{'}s University (NWU), Nara, Japan\\
$^{77}$ National and Kapodistrian University of Athens, School of Science, Department of Physics , Athens, Greece\\
$^{78}$ National Centre for Nuclear Research, Warsaw, Poland\\
$^{79}$ National Institute of Science Education and Research, Homi Bhabha National Institute, Jatni, India\\
$^{80}$ National Nuclear Research Center, Baku, Azerbaijan\\
$^{81}$ National Research and Innovation Agency - BRIN, Jakarta, Indonesia\\
$^{82}$ Niels Bohr Institute, University of Copenhagen, Copenhagen, Denmark\\
$^{83}$ Nikhef, National institute for subatomic physics, Amsterdam, Netherlands\\
$^{84}$ Nuclear Physics Group, STFC Daresbury Laboratory, Daresbury, United Kingdom\\
$^{85}$ Nuclear Physics Institute of the Czech Academy of Sciences, Husinec-\v{R}e\v{z}, Czech Republic\\
$^{86}$ Oak Ridge National Laboratory, Oak Ridge, Tennessee, United States\\
$^{87}$ Ohio State University, Columbus, Ohio, United States\\
$^{88}$ Physics department, Faculty of science, University of Zagreb, Zagreb, Croatia\\
$^{89}$ Physics Department, Panjab University, Chandigarh, India\\
$^{90}$ Physics Department, University of Jammu, Jammu, India\\
$^{91}$ Physics Department, University of Rajasthan, Jaipur, India\\
$^{92}$ Physics Program and International Institute for Sustainability with Knotted Chiral Meta Matter (SKCM2), Hiroshima University, Hiroshima, Japan\\
$^{93}$ Physikalisches Institut, Eberhard-Karls-Universit\"{a}t T\"{u}bingen, T\"{u}bingen, Germany\\
$^{94}$ Physikalisches Institut, Ruprecht-Karls-Universit\"{a}t Heidelberg, Heidelberg, Germany\\
$^{95}$ Physik Department, Technische Universit\"{a}t M\"{u}nchen, Munich, Germany\\
$^{96}$ Politecnico di Bari and Sezione INFN, Bari, Italy\\
$^{97}$ Research Division and ExtreMe Matter Institute EMMI, GSI Helmholtzzentrum f\"ur Schwerionenforschung GmbH, Darmstadt, Germany\\
$^{98}$ Saha Institute of Nuclear Physics, Homi Bhabha National Institute, Kolkata, India\\
$^{99}$ School of Physics and Astronomy, University of Birmingham, Birmingham, United Kingdom\\
$^{100}$ Secci\'{o}n F\'{\i}sica, Departamento de Ciencias, Pontificia Universidad Cat\'{o}lica del Per\'{u}, Lima, Peru\\
$^{101}$ Stefan Meyer Institut f\"{u}r Subatomare Physik (SMI), Vienna, Austria\\
$^{102}$ SUBATECH, IMT Atlantique, Nantes Universit\'{e}, CNRS-IN2P3, Nantes, France\\
$^{103}$ Suranaree University of Technology, Nakhon Ratchasima, Thailand\\
$^{104}$ Technical University of Ko\v{s}ice, Ko\v{s}ice, Slovak Republic\\
$^{105}$ The Henryk Niewodniczanski Institute of Nuclear Physics, Polish Academy of Sciences, Cracow, Poland\\
$^{106}$ The University of Texas at Austin, Austin, Texas, United States\\
$^{107}$ Universidad Aut\'{o}noma de Sinaloa, Culiac\'{a}n, Mexico\\
$^{108}$ Universidade de S\~{a}o Paulo (USP), S\~{a}o Paulo, Brazil\\
$^{109}$ Universidade Estadual de Campinas (UNICAMP), Campinas, Brazil\\
$^{110}$ Universidade Federal do ABC, Santo Andre, Brazil\\
$^{111}$ University of Cape Town, Cape Town, South Africa\\
$^{112}$ University of Houston, Houston, Texas, United States\\
$^{113}$ University of Jyv\"{a}skyl\"{a}, Jyv\"{a}skyl\"{a}, Finland\\
$^{114}$ University of Kansas, Lawrence, Kansas, United States\\
$^{115}$ University of Liverpool, Liverpool, United Kingdom\\
$^{116}$ University of Science and Technology of China, Hefei, China\\
$^{117}$ University of South-Eastern Norway, Kongsberg, Norway\\
$^{118}$ University of Tennessee, Knoxville, Tennessee, United States\\
$^{119}$ University of the Witwatersrand, Johannesburg, South Africa\\
$^{120}$ University of Tokyo, Tokyo, Japan\\
$^{121}$ University of Tsukuba, Tsukuba, Japan\\
$^{122}$ University Politehnica of Bucharest, Bucharest, Romania\\
$^{123}$ Universit\'{e} Clermont Auvergne, CNRS/IN2P3, LPC, Clermont-Ferrand, France\\
$^{124}$ Universit\'{e} de Lyon, CNRS/IN2P3, Institut de Physique des 2 Infinis de Lyon, Lyon, France\\
$^{125}$ Universit\'{e} de Strasbourg, CNRS, IPHC UMR 7178, F-67000 Strasbourg, France, Strasbourg, France\\
$^{126}$ Universit\'{e} Paris-Saclay, Centre d'Etudes de Saclay (CEA), IRFU, D\'{e}partment de Physique Nucl\'{e}aire (DPhN), Saclay, France\\
$^{127}$ Universit\'{e}  Paris-Saclay, CNRS/IN2P3, IJCLab, Orsay, France\\
$^{128}$ Universit\`{a} degli Studi di Foggia, Foggia, Italy\\
$^{129}$ Universit\`{a} del Piemonte Orientale, Vercelli, Italy\\
$^{130}$ Universit\`{a} di Brescia, Brescia, Italy\\
$^{131}$ Variable Energy Cyclotron Centre, Homi Bhabha National Institute, Kolkata, India\\
$^{132}$ Warsaw University of Technology, Warsaw, Poland\\
$^{133}$ Wayne State University, Detroit, Michigan, United States\\
$^{134}$ Westf\"{a}lische Wilhelms-Universit\"{a}t M\"{u}nster, Institut f\"{u}r Kernphysik, M\"{u}nster, Germany\\
$^{135}$ Wigner Research Centre for Physics, Budapest, Hungary\\
$^{136}$ Yale University, New Haven, Connecticut, United States\\
$^{137}$ Yonsei University, Seoul, Republic of Korea\\
$^{138}$  Zentrum  f\"{u}r Technologie und Transfer (ZTT), Worms, Germany\\
$^{139}$ Affiliated with an institute covered by a cooperation agreement with CERN\\
$^{140}$ Affiliated with an international laboratory covered by a cooperation agreement with CERN.\\

\end{flushleft} 

\end{document}